# H(II) CENTERS IN NATURAL SILICA UNDER REPEATED UV LASER IRRADIATIONS

F. Messina *, M. Cannas, R. Boscaino

*INFM and Dipartimento di Scienze Fisiche ed Astronomiche, Università di Palermo,*

*via Archirafi 36, I-90123, Palermo, Italy*

**Abstract:**

We investigated the kinetics of H(II) centers (=Ge$^\bullet$-H) in natural silica under repeated 266nm UV irradiations performed by a Nd:YAG pulsed laser. UV photons temporarily destroy these paramagnetic defects, their reduction being complete within 250 pulses. After re-irradiation, H(II) centers grow again, and the observed recovery kinetics depends on the irradiation dose; multiple 2000 pulses re-irradiations induce the same post-irradiation kinetics of H(II) centers after each exposure cycle. The analysis of these effects allows us to achieve a deeper understanding of the dynamics of the centers during and after laser irradiation.

*Corresponding author: F. Messina

INFM and Dip.to di Scienze Fisiche ed Astronomiche, via Archirafi 36, I-90123 Palermo.

Phone: +39 0916234218, Fax: +390 916162461, e-mail: fmessina@fisica.unipa.it

# 1. Introduction

In technological applications of Ge-containing silica glasses, the presence of hydrogen appears to be particularly relevant for its tendency to enhance the photosensitivity of the materials used to produce Bragg gratings in fibers [1-3]. In fact, it has been hypothesized that a contribution to photosensitivity arises from conversion processes of Ge-related defects, one of which is the transformation of the optically active twofold coordinated Ge (=Ge$^{\bullet\bullet}$) into the paramagnetic H(II) center (=Ge$^{\bullet}$-H) by trapping a H$_0$ atom [2, 4-6]:

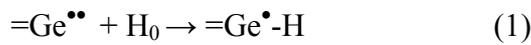

$$=Ge^{\bullet\bullet} + H_0 \rightarrow =Ge^{\bullet}\text{-H} \qquad (1)$$

where (=) stands for bonds with two oxygen atoms, and each ($^{\bullet}$) stands for an electron.

Natural silica samples contain a small amount (~1ppm) of Ge impurities, mainly in twofold coordinated form, due to natural contamination [7]. In these materials, at room temperature, reaction (1) constitutes the main Ge-related conversion process stimulated by UV radiation and the growth of H(II) mainly occurs in the post-irradiation stage, rather than during it [8,9]. Hydrogen available for the formation of H(II) results from the following steps: during irradiation, radiolytic H$_0$ from SiH or OH bonds dimerizes in H$_2$ [10]; then, H$_0$ for reaction (1) is made available in the post-irradiation stage by breaking of H$_2$ at paramagnetic centers [5], such as E' centers (≡Si$^{\bullet}$) [11]. Hence, the growth kinetics of H(II) centers is mainly controlled by the diffusion parameters of H$_2$.

In the study of the reaction dynamics giving rise to the formation of H(II) centers, the direct effect of UV radiation on these defects needs to be clarified. To this aim, in this work we have investigated the growth of H(II) centers under repeated UV laser irradiations. The characteristic feature of this experimental approach is the observation of the effects induced by UV light on a previously irradiated sample, which differs from an as-grown material as regards the typologies of present defects. It is worth to note that the current results may also be helpful to understand the response of photosensitive silica-based devices to repeated UV exposures.

## 2. Materials and Methods

Two natural silica materials supplied by Heraeus QuartzGlas where employed in our experiments: a type I dry Infrasil 301, OH content ~10ppm, and a type II wet Herasil 1, OH content ~150ppm. The samples were slab shaped with sizes of 5x5x1mm$^3$. UV exposure with 266nm photons was performed at room temperature using the fourth harmonic generation from the pulsed radiation of a Nd:YAG laser (Quanta System SYL 201), at a repetition rate of 1 Hz, each pulse having energy density of W=40mJ/cm$^2$ and 5ns duration.

Post irradiation kinetics of the UV-induced paramagnetic centers were investigated at room temperature by electron spin resonance (ESR) measurements, performed at different times (from $10^2$s to $10^6$s) after laser exposure on a spectrometer (Bruker EMX) working at 9.8GHz. The ESR signature of H(II) center consists in a doublet with a split of 11.8mT due to the hyperfine interaction between the unpaired electronic spin on Ge and the nuclear one of H [4,5]. This signal was detected with microwave power P=3.2mW, low enough to prevent saturation, and a 100 kHz modulation field of peak-to-peak amplitude $B_m$=0.4mT. The concentration of paramagnetic H(II) centers was evaluated by comparing the double-integrated ESR spectra with that of E' centers [11, 12], whose absolute density was determined with accuracy of ±20% by spin-echo measurements [13].

## 3. Results

As already put forward in a previous paper [8], 266nm Nd:YAG laser irradiation of natural silica induces the formation of H(II) centers mostly occurring in the post-irradiation stage. In Fig. 1 is shown the dose dependence of the defect concentration [H(II)], as measured in wet silica samples. Open symbols represent concentrations measured just after the end of irradiation (hereafter called initial concentrations), whereas full symbols are the stationary concentrations, measured many days after the end of irradiation, when the diffusion-limited growth is completed [8]. Typical growth kinetics from initial concentration to stationary concentration is shown in the inset, which refers to the sample irradiated with 2000 laser pulses. The stationary concentration grows with the number of pulses and after ~300 shots it saturates to $(2.1\pm0.2)10^{15}cm^{-3}$. We stress that the results of Fig. 1 are representative of all natural silica samples investigated.

To study the effect of UV radiation on H(II) centers we started from samples preliminary exposed to an irradiation dose high enough (2000 shots) to produce the maximum H(II) concentration. We re-irradiated a Herasil 1 sample with an increasing number of laser shots, and between successive exposures we measured the defect concentration. We found that, after each exposure, the defect concentration remains almost stationary (within ~10%) during the time interval required for measurement (~500s). In Fig. 2 variations of [H(II)] observed during the exposure cycle are shown. It is apparent that the new exposure (which we will call re-irradiation) results in the destruction of H(II) centers which had grown after the first high-dose irradiation, their concentration decreasing to $(4.5\pm0.5)10^{14}cm^{-3}$, i.e. ~25% of the initial value, after ~250 laser pulses. The H(II) reduction with the number of pulses N is well fitted by an exponential law:

$$[H(II)]=A\times exp(-N/N_0)+cost \qquad (2)$$

where $N_0=30\pm3$, the correlation coefficient being R=0.995. The same reduction effect was evidenced also in natural dry $SiO_2$ as shown in the inset of Fig. 2.

After the whole 250 pulses sequence, [H(II)] grows up again but does not recover its initial value within the investigated time scale: their concentration increases to $(1.4\pm0.1)\cdot10^{15}cm^{-3}$ in $2\times10^6$ s. Our study was completed investigating the recovery kinetics observed after a longer re-irradiation. Specifically, we performed up to 4 identical high-dose exposures (2000 pulses) on another sample, and after each irradiation we measured the post-irradiation kinetics of H(II) centers until completion. Results of this experiment are reported in Fig. 3. The first irradiation induces the growth of H(II) centers from an initial value of $(5.0\pm0.5)10^{14}cm^{-3}$, measured after $10^2$s to $(1.8\pm0.2)10^{15}cm^{-3}$, measured after $5\times10^5$s. Just after the second UV exposure, the H(II) centers concentration is reduced again to $(5.0\pm0.5)10^{14}cm^{-3}$, after which it increases nearly with the same kinetics observed after the previous irradiation and reaches again the value of $(1.8\pm0.2)10^{15}cm^{-3}$. The same results, shown in panels (c) and (d) are observed also after a third and a fourth irradiation. For comparison, in panel (b) the kinetics observed after the 250 pulses sequence is also shown.

## 4. Discussion

One of the main evidences arising from our data is that H(II) centers, which are indirectly generated by reaction (1) activated by irradiation, are also reduced by laser exposure. Since the same photo-induced destruction effect is observed both in wet and dry natural samples, we hypothesize that it is caused by the direct absorption of hv=4.66 eV light at the defect site regardless of the material type. From the $N_0$ value measured by exponential fit of data in Fig. 2, we calculate the cross section for the photo-induced destruction of H(II) center: $\sigma=N_0^{-1}(h\nu/W)=(6.2\pm0.6)\times10^{-19}cm^2$. Then, the H(II) concentration, $c=(2.1\pm0.2)10^{15}cm^{-3}$, leads to the mimimum value of the 4.66 eV absorption coefficient of H(II) centers, $\alpha=\sigma c=(1.3\pm0.2)\cdot10^{-3}cm^{-1}$, if we suppose that each absorbed photon results in the destruction of a defect. We observe that our sensitivity limit for optical absorption measurements in these samples is $10^{-2}$ cm$^{-1}$ [8], preventing us from detecting the anticipated absorption signal in this region which could be ascribed to H(II) centers.

The observation of the photo-induced decay of H(II) centers allows us to understand the peculiar feature of the growth kinetics of these defects, that are formed mostly in the post-irradiation stage, rather than during it, as apparent from data in Fig. 1. In fact, the stationary concentrations result 3-4 times higher than initial concentrations, regardless of exposure duration. This result is explained as follows: formation of H(II) centers during irradiation through reaction (1) is inhibited because of the competition with the photo-induced decay of the centers.

Now we address the post-irradiation kinetics observed after a sequence of high-dose (2000 pulses) laser irradiations. The results evidence that the growth of H(II) centers repeats itself with the same kinetics after each exposure cycle, via two steps:

1. Each re-irradiation temporarily destroys H(II) centers and reduces their concentration from $(1.8\pm0.2)\ 10^{15} cm^{-3}$ back to $(4.5\pm0.5)10^{-14} cm^{-3}$. It is worth to note that this reduction is completed just after the first 250 shots, as visible in Fig. 2.

2. After each re-irradiation, the sample loses memory of its previous history, meaning that all kinetics repeat themselves identically to the first. This effect is peculiar of high dose re-irradiations, as emerges from comparison with the post-irradiation kinetics observed after 250 shots.

On the basis of the repeatability of the H(II) kinetics, we infer that the decay of these centers leads to recovery of their precursor, the twofold coordinated Ge, which may occur through the following photo-reaction:

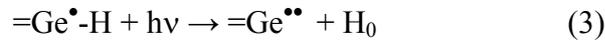
$$=Ge^{\bullet}\text{-}H + h\nu \rightarrow =Ge^{\bullet\bullet} + H_0 \qquad (3)$$

In principle, the post-irradiation kinetics is determined by the initial concentrations of the centers taking part to the reactions which induce the growth of H(II). From this point of view, the memory loss evidenced in Fig. 3 implies that each high dose re-irradiation is able to reset the concentrations of the centers to the same values they had after the first exposure. Apart from $=Ge^{\bullet\bullet}$, we recall the fundamental role played by the E' centers. In fact, the atomic hydrogen needed to form H(II) center by reaction (1) is made available by the breaking of $H_2$ at E' centers [8, 14]:

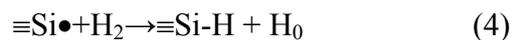
$$\equiv Si\bullet + H_2 \rightarrow \equiv Si\text{-}H + H_0 \qquad (4)$$

Consistently, in natural silica also the E' centers repeat the same annealing kinetics many times after each high dose re-irradiation [14]. The combination of the two results suggests that in a long exposure the concentrations of E', H(II) and hydrogen reach some equilibrium values regardless of the previous history of the sample, so leading to repeatability of the post-irradiation kinetics.

As a final remark, we acknowledge that in Ref. [15], Radzig et al. observed the photo-induced decay of H(I) center (=Si$^\bullet$-H). Their experimental observation is connected with the present result, since H(I) and H(II) both belong to an isoelectronic series of defects, localized respectively on a Si, Ge, Sn atom, which are known to have similar formation and spectroscopic properties [15-17]. From this point of view, our data suggest that the isoelectronic defects share also the same photo-induced decay properties.

## 5. Conclusions

UV radiation at 266nm causes the destruction of H(II) centers (=Ge$^\bullet$-H), supposedly by back-conversion to their precursor =Ge$^{\bullet\bullet}$. Multiple high-dose irradiations result in the repetition of the same post-irradiation kinetics of H(II) centers after each exposure. This effect is achieved by the photo-decomposition of previously formed defects and by the ability of each high dose exposure to reset to fixed values the concentrations of the centers involved in the post-irradiation phenomena.

## Acknowledgements

The authors thank S. Agnello, G. Buscarino and F.M. Gelardi for useful discussions and G. Lapis and G. Napoli for technical assistance. This work is part of a national project (PRIN2002) supported by the Italian Ministry of University Research and Technology.

**Figure captions:**

**Figure 1**) Laser dose dependence of H(II) centers concentration observed in wet natural $SiO_2$. Open symbols are the concentrations just after the end of irradiation, full symbols are the stationary concentrations, measured many days after the end of UV exposure. Inset shows the typical growth kinetics of H(II) from initial to stationary concentration, as observed in the sample irradiated with 2000 pulses.

**Figure 2**) Variations of H(II) centers concentration induced by re-irradiation of a natural wet sample preliminary exposed to 2000 laser shots; solid line plots the exponential best fit of the data. The inset shows the results obtained in natural dry $SiO_2$.

**Figure 3**) H(II) kinetics induced in natural $SiO_2$ by a cycle of 4 repeated laser exposures, 2000 shots for each one. Open symbols in the second graph refer to the growth kinetics observed after the 250 shots re-irradiation of Fig.2.

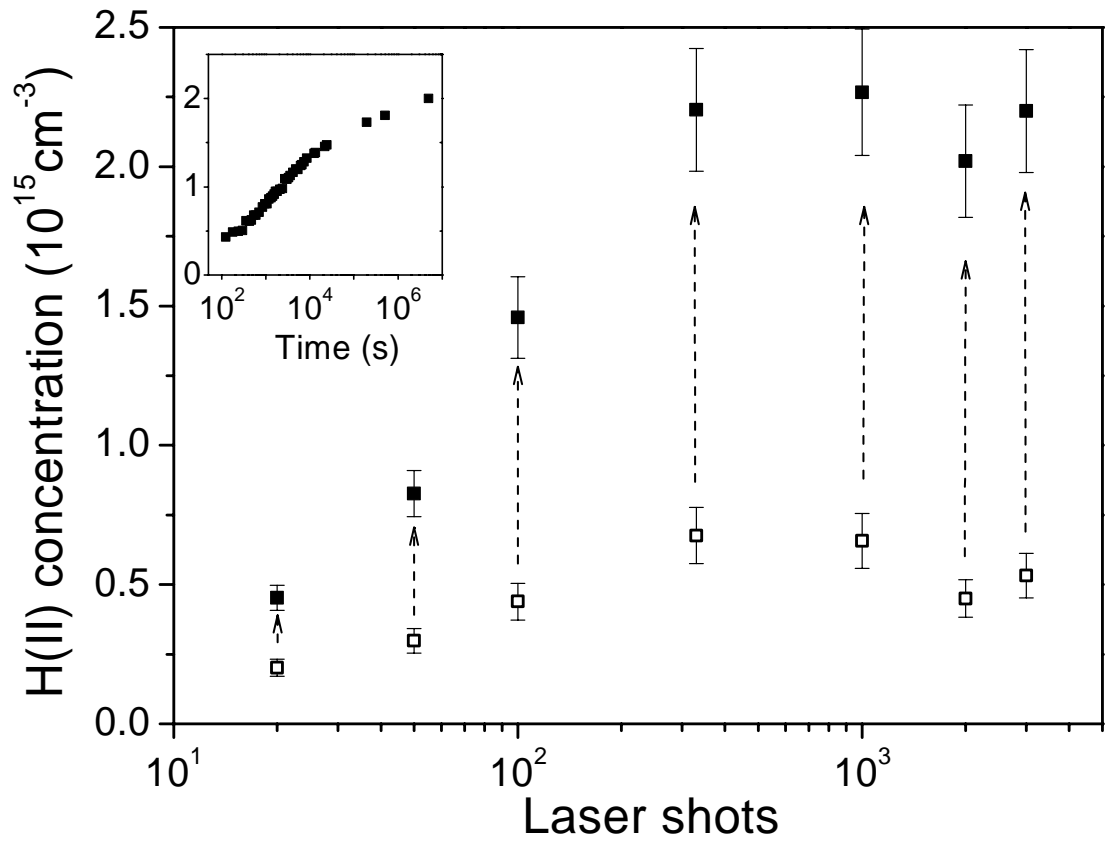

**Figure 1**

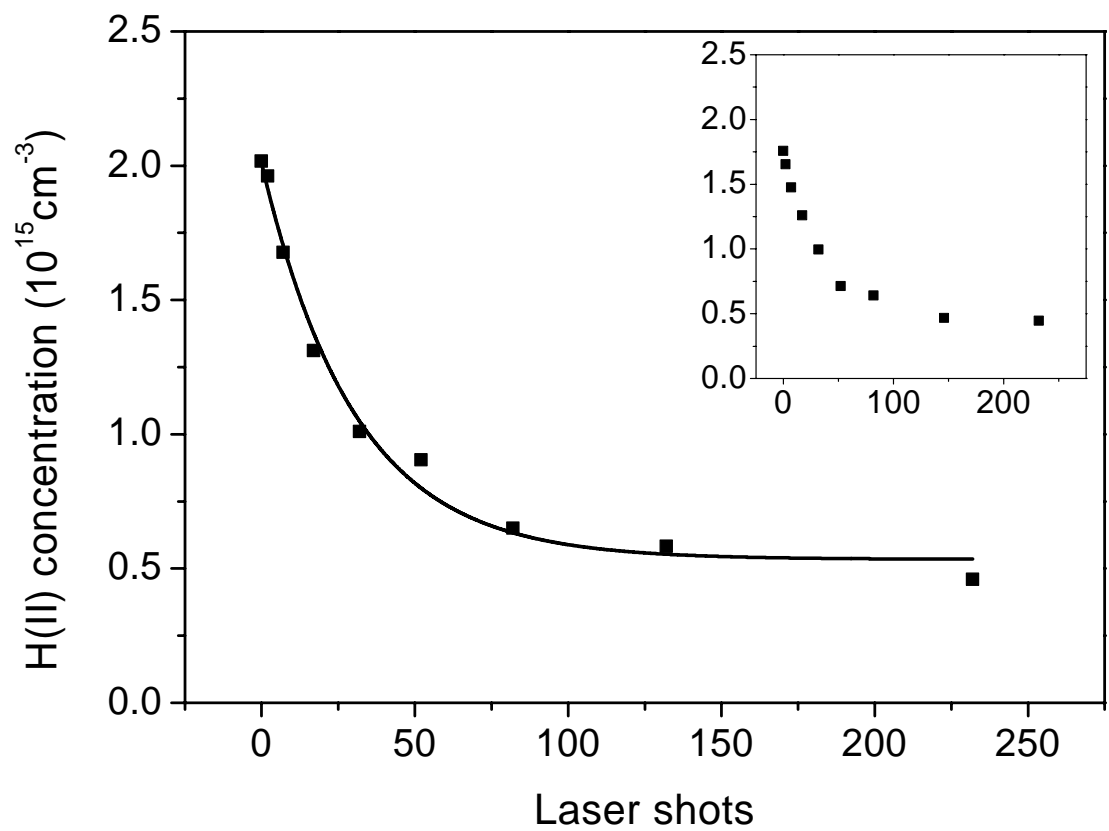

**Figure 2**

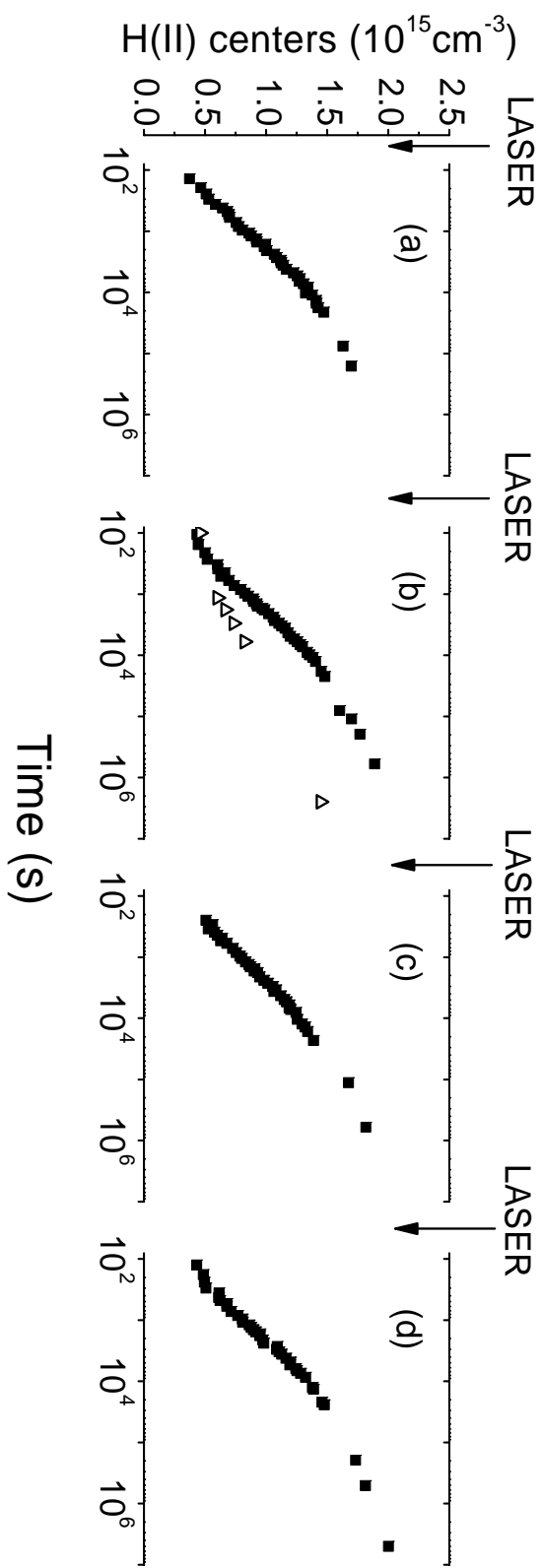

**Figure 3**